\documentclass[oldversion,longauth]{aa}

\def\simlt{\lower.5ex\hbox{$\; \buildrel < \over \sim \;$}}
\def\simgt{\lower.5ex\hbox{$\; \buildrel > \over \sim \;$}}
\usepackage{subfigure}
\usepackage[dvips]{graphicx}
\usepackage[latin1]{inputenc}
\usepackage{amsmath}
\usepackage{amssymb}
\usepackage{verbatim}

\begin{document}

\title{zCOSMOS 10k-bright spectroscopic sample
\thanks{based on data obtained with the European Southern Observatory
Very Large Telescope, Paranal, Chile, program 175.A-0839}
}
\subtitle{Exploring mass and environment dependence \\in early-type galaxies}

\author{
      M.~Moresco \inst{1} 
\and L.~Pozzetti    \inst{2}  
\and A.~Cimatti\inst{1} 
\and G.~Zamorani \inst{2} 
\and M.~Mignoli \inst{2}
\and S. Di Cesare  \inst{1}
\and M.~Bolzonella  \inst{2}  
\and E.~Zucca \inst{2} 
\and S.~Lilly \inst{3}
\and K.~Kova\v{c}\inst{3} 
\and M.~Scodeggio\inst{4} 
\and P.~Cassata \inst{5,6} 
\and L.~Tasca \inst{4,6} 
\and D.~Vergani \inst{2} 
\and C. Halliday\inst{7} 
\and M.~Carollo \inst{3} 
\and T.~Contini\inst{8} 
\and J.-P.~Kneib\inst{6} 
\and O.~Le~F\'evre\inst{6} 
\and V.~Mainieri\inst{9}
\and A.~Renzini\inst{10} 
\and S.~Bardelli\inst{2} 
\and A.~Bongiorno\inst{9} 
\and K.~Caputi\inst{11} 
\and G.~Coppa\inst{2} 
\and O.~Cucciati\inst{6,12} 
\and S.~de~la~Torre\inst{6} 
\and L.~de~Ravel\inst{6} 
\and P.~Franzetti\inst{4} 
\and B.~Garilli\inst{4} 
\and A.~Iovino\inst{12} 
\and P.~Kampczyk\inst{3} 
\and C.~Knobel\inst{3} 
\and F.~Lamareille \inst{8} 
\and J.-F.~Le~Borgne\inst{8} 
\and V.~Le~Brun\inst{7} 
\and C.~Maier\inst{3}
\and R.~Pell\`o\inst{8} 
\and Y.~Peng \inst{3} 
\and E.~Perez~Montero\inst{8}
\and E.~Ricciardelli\inst{10} 
\and J.D.~Silverman\inst{3}
\and M.~Tanaka\inst{9}
\and L.~Tresse\inst{6} 
\and U.~Abbas\inst{6} 
\and D.~Bottini\inst{4} 
\and A.~Cappi\inst{2} 
\and L.~Guzzo\inst{12} 
\and A.M.~Koekemoer\inst{13} 
\and A.~Leauthaud\inst{6}
\and D.~Maccagni\inst{4} 
\and C.~Marinoni\inst{14} 
\and H.J.~McCracken\inst{15} 
\and P.~Memeo\inst{4} 
\and B.~Meneux\inst{8,16} 
\and P.~Nair\inst{2} 
\and P.~Oesch\inst{3}
\and C.~Porciani\inst{3}
\and R.~Scaramella\inst{17} 
\and C.~Scarlata\inst{18} 
\and N.~Scoville\inst{19} 
}

   \offprints{Michele Moresco (\email{michele.moresco@unibo.it}) }

\institute{
Dipartimento di Astronomia, Universit\'a degli Studi di Bologna, via Ranzani 1, I-40127, Bologna, Italy
\and 
INAF -- Osservatorio Astronomico di Bologna, via Ranzani 1, I-40127, Bologna, Italy
\and 
Institute of Astronomy, Swiss Federal Institute of Technology (ETH H\"onggerberg), CH-8093, Z\"urich, Switzerland
\and 
INAF -- Istituto di Astrofisica Spaziale e Fisica Cosmica di Milano, via Bassini 15, I-20133 Milano, Italy;
\and 
Department of Astronomy, University of Massachusetts, 710 North Pleasant Street, Amherst, MA 01003, USA
\and 
{Laboratoire d'Astrophysique de Marseille, Universit\'{e} d'Aix-Marseille, CNRS, 38 rue Frederic Joliot-Curie, F-13388 Marseille Cedex 13, France}
\and 
{INAF -- Osservatorio Astrofisico di Arcetri, Largo Enrico Fermi 5, I-50125 Firenze, Italy }
\and 
Laboratoire d'Astrophysique de Toulouse-Tarbes, Universit\'{e} de Toulouse, CNRS, 14 avenue Edouard Belin, F-31400 Toulouse, France
\and 
Max Planck Institut f\"ur Extraterrestrische Physik,  Giessenbachstrasse, D-84571 Garching, Germany
\and 
Dipartimento di Astronomia, Universit\'a di Padova, Padova, Italy
\and 
{SUPA Institute for Astronomy, The University of Edinburgh, Royal Observatory, Blackford Hill, Edinburgh, EH9 3HJ, United Kingdom}
\and 
INAF -- Osservatorio Astronomico di Brera, via Brera 28, I-20121  Milano, Italy
\and 
{Space Telescope Science Institute, 3700 San Martin Drive, Baltimore, MD 21218, USA}
\and 
Centre de Physique Theorique, Marseille, Marseille, France
\and 
Institut d'Astrophysique de Paris, Universit\'e Pierre \& Marie Curie, Paris, France
\and 
Universitats Sternwarte, Scheinerstrasse 1, D-81679 Muenchen
\and 
{INAF -- Osservatorio Astronomico di Roma, via di Frascati 33, I-00040 Monteporzio Catone, Italy}
\and 
Spitzer Science Center, Pasadena, CA, USA
\and 
{California Institute of Technology, MC 105-24, 1200 East California Boulevard, Pasadena, CA 91125, USA}
}

\authorrunning {M. Moresco, et al.}

\titlerunning {Dependence on mass and environment of zCOSMOS ETGs}

\date{Received -- -- ----; accepted -- -- ----}

\abstract{
{\it Aims.} We present the analysis of the U-V rest-frame color distribution and some spectral features as a function of mass and environment for a sample of early-type galaxies up to $z=1$ extracted from the zCOSMOS spectroscopic survey. This analysis is used to place constraints on the relative importance of these two properties in controlling galaxy evolution.\\
{\it Methods.} We used the zCOSMOS 10k-bright sample, limited to the AB magnitude range $15<I<22.5$, from which we extracted two different subsamples of early-type galaxies. The first sample ("{\it red galaxies}") was selected using a photometric classification (2098 galaxies), while in the second case ("{\it ETGs}") we combined morphological, photometric, and spectroscopic properties to obtain a more reliable sample of elliptical, red, passive, early-type galaxies (981 galaxies). The analysis is performed at fixed mass to search for any dependence of the color distribution on environment, and at fixed environment to search for any mass dependence.\\
{\it Results.} In agreement with the low redshift results of the SDSS, we find that the color distribution of {\it red galaxies} is not strongly dependent on environment for all mass bins, exhibiting only a weak trend such that galaxies in overdense regions ($log_{10}(1+\delta)\sim1.2$) are redder than galaxies in underdense regions ($log_{10}(1+\delta)\sim0.1$), with a difference of $\left<\Delta(U-V)_{rest}\right>=0.027\pm0.008$ mag. On the other hand, the dependence on mass is far more significant, and we find that the average colors of massive galaxies ($log_{10}(M/M_{\odot})\sim10.8$) are redder by $\left<\Delta(U-V)_{rest}\right>=0.093\pm0.007$ mag than low-mass galaxies ($log_{10}(M/M_{\odot})\sim10$) throughout the entire redshift range. We study the color-mass $(U-V)_{rest}\propto S_{M}\cdot log_{10}(M/M_{\odot})$ relation, finding a mean slope $\left<S_{M}\right>=0.12\pm0.005$, while the color-environment $(U-V)_{rest}\propto S_{\delta}\cdot log_{10}(1+\delta)$ relation is flatter, with a slope always smaller than $S_{\delta}\approx0.04$.\\
The spectral analysis that we perform on our {\it ETGs} sample is in good agreement with our photometric results: we study the 4000 \AA break and the equivalent width of the $H\delta$ Balmer line, finding for D4000 a dependence on mass ($\left<\Delta D4000\right>=0.11\pm0.02$ between $log_{10}(M/M_{\odot})\sim10.2$ and $log_{10}(M/M_{\odot})\sim10$.8), and a much weaker dependence on environment ($\left<\Delta D4000\right>=0.05\pm0.02$ between high and low environment quartiles). The same is true for the equivalent width of $H\delta$, for which we measure a difference of $\Delta EW_{0}(H\delta)=0.28\pm0.08$\AA across the same mass range and no significant dependence on environment.
By analyzing the lookback time of early-type galaxies, we support the possibility of a downsizing scenario, in which massive galaxies with a stronger D4000 and an almost constant equivalent width of $H\delta$ formed their mass at higher redshift than lower mass ones.\\
We also conclude that the main driver of galaxy evolution is the galaxy mass, the environment playing a subdominant role.

         }

\maketitle

%-------------------------------------------------------------------------------
\section{Introduction}\label{sec:intro}
Understanding the evolution of early-type galaxies (ETGs) is one of the most interesting and intensely studied topics in modern astronomy.
A fundamental step in determining how galaxies have assembled their mass and evolved is to study in detail the properties of galaxies as a function of redshift (see Renzini 2006). In this respect, the use of galaxy colors has many advantages. On the one hand, they are directly linked to observables, unlike, for example, mass, age, and star formation rate. On the other hand, it has been proven that the color distribution exhibits a bimodality that corresponds to a division between late-type (younger, star-forming, bluer) and early-type (older, passive, redder) galaxies; it is therefore possible to use colors instead of morphological types to classify and select the galaxy population, and also to use the colors of early-type galaxies as a proxy of their age.\\
Considering early-type galaxies, many studies of their properties have been performed to search for correlations between different observables, such as the fundamental plane, the color-magnitude relation, and the luminosity-velocity dispersion relation. In this respect, it is important to quantify the strength of the dependences as a function of mass, luminosity, and environment. Low redshift results indicate that the mean of and the variance in the color distribution are strong functions of either luminosity or stellar mass (Bernardi et al. 2003a, Bernardi et al. 2003c, Baldry et al. 2004), and  similar trends are observed out to $z\sim1$ (Bell et al. 2004). On the  other hand, the dependence on environment seems to be weaker (Balogh et al. 2004, Hogg et al. 2004, Bernardi et al. 2005). Massive early-type galaxies in low-density environments are on average younger and slightly more metal-rich than their counterparts in high-density environments (see Thomas et al. 2005, Bernardi et al. 2006, Gallazzi et al. 2006, Chang et al. 2006, Kurk et al. 2009). However, the mean color seems to be only weakly correlated with environment (Balogh et al. 2004) and Bolzonella et al. (2009) found that the shape of the galaxy stellar mass functions of early-type and late-type galaxies is similar in the high and low environments probed by the zCOSMOS survey.\\
In this paper, we present a study of the $(U-V)_{rest}$ color for two different samples of ETGs ($\sim 2100$ and $\sim 1000$ galaxies, respectively) extracted from the 10-k zCOSMOS spectroscopic data set in the range $0.1 < z \leq 1$ (see Lilly et al. (2009) for a detailed description of the sample). We analyze their color distributions, investigating the dependence of their shapes on mass and environment.\\
In addition to colors, we decided to analyze two spectroscopic line indices in galaxy spectra, the 4000 \AA break (D4000) and the equivalent width of the $H\delta$ Balmer line ($EW_{0}(H\delta)$), which have proven to be good age indicators (see Hamilton 1985, Bruzual \& Charlot 1993, Kauffmann et al. 2003). The D4000 index is produced by the blending of a large number of spectral lines in a narrow wavelength region, which creates a break in the spectrum at this wavelength. The main contribution to the opacity of a star comes from ionized metals: in hot stars, the metals are multiply ionized and the opacity decreases, so the D4000 will be small for young stellar populations and large for old, metal-rich galaxies. It is therefore an observable that is strongly dependent on the age and metallicity of a galaxy. On the other side, the $EW_{0}(H\delta)$ correlates well with recent episodes of star formation. To verify what we found from the analysis of colors, we also study these spectral line strengths as a function of redshift in different mass and environment bins.
Finally, we evaluate the lookback time of {\it ETGs} and find that more massive galaxies have assembled their mass earlier than less massive galaxies, showing a trend in agreement with the one found in the analysis of low redshift galaxies and supporting a downsizing scenario.\\
The paper is organized in the following way. In Sect. \ref{sec:data}, we present the zCOSMOS data sample, describe the different criteria with which ETGs have been selected, and give the definitions of environment. In Sect. \ref{sec:analysis}, we describe the mass cuts and the adopted definitions for D4000 and $EW_{0}(H\delta)$; we also discuss the selection bias that may be present in our data.
In Sect. \ref{sec:res} we present our results on the analysis of the ETG samples by separately assessing the contributions of mass and environment. We show the trends in mass and environment of ETGs found from both a photometric and a spectral point of view, studying the slopes of the color-mass and color-environment relations. Finally, we describe how we analyzed the dependence of the lookback time of formation of ETGs on stellar mass, supporting a downsizing scenario.

Throughout the paper, we adopt the cosmological parameters $H_{0}=70$ $km$ $s^{-1}$ $Mpc^{-1}$, $\Omega_{m}=0.25$, $\Omega_{\Lambda}=0.75$. Magnitudes are quoted in the AB system.

%-------------------------------------------------------------------------------
\section{Data}\label{sec:data}

The COSMOS Survey (Scoville et al. 2007) is the largest contiguous HST survey ever undertaken ($\approx640$ orbits, Koekemoer et al. 2007), which has imaged a $\sim 2$ deg$^2$ field using the Advanced Camera for Surveys (ACS) with single-orbit I-band exposures to a depth $I_{AB}\simeq28$ mag and $50\%$ completeness at $I_{AB}=26.0$ mag for sources $0.5''$ in diameter. COSMOS observations are implemented with an excellent coverage of the field with multiband photometry from the UV (with GALEX, Zamojski et al. 2007), optical (with Subaru and CFHT, Taniguchi et al. 2007, Capak et al. 2007), NIR (with CTIO, KPNO, Capak et al. 2007, and CFHT, McCracken et al. 2010), to MIR and FIR (with Spitzer, Sanders et al. 2007), in combination with a multiwavelength dataset from radio (with VLA, Schinnerer et al. 2007), millimeter (with MAMBO, Bertoldi et al. 2007), to X-rays (with XMM, Hasinger et al. 2007, and Chandra, Elvis et al. 2009).

The analysis presented in this paper is based on the zCOSMOS spectroscopic survey (Lilly et al. 2007). This is an ongoing ESO Large Programme
($\sim 600$ hours of observations) aiming to map the COSMOS field with the VIsible Multi-Object Spectrograph (VIMOS, Le~F\`evre et al.), mounted on the ESO Very Large Telescope (VLT). We use the {\it bright} part of zCOSMOS survey, which consists of spectroscopy limited to objects in the magnitude range $15.0<I<22.5$. A medium resolution grism ($R\approx600$) was used with a slit width of 1 arcsec, to achieve a velocity accuracy of $\sim100 km/s$ and enable redshifts to be measured with a high success rate using one hour integrations. The spectral ranges of the observations are typically $5550-9650${\AA}. For more details about the zCOSMOS 10k-bright sample that we use in this analysis, we refer to Lilly et al. (2009).\\
Among the multi-band photometric data of the COSMOS field, we used the observed magnitudes in 10 photometric bands (CFHT $u^*$ and $K_s$, Subaru $B_J$, $V_J$, $g^+$, $r^+$, $i^+$, and $z^+$, and Spitzer IRAC at 3.6 $\mu$m and 4.5 $\mu$m). The photometric catalog is described in Capak et al. (2007). Following their approach, magnitudes were corrected for Galactic extinction using Table 11 of Capak et al. (2007) and the photometry was optimized by applying zeropoint offsets to the observed magnitudes to reduce differences between the observed and reference magnitudes computed from a set of template spectral energy distributions (SEDs). From the total sample of $\sim$10 000 objects with measured redshifts, we used objects within the statistical sample defined in the magnitude range $15 <I< 22.5$, and removed the spectroscopically confirmed stars, the broad-line AGNs, as well as the galaxies with low quality redshift flags (Lilly et al. 2009, Bolzonella et al. 2009, Pozzetti et al. 2009).

\subsection {ETG selection criteria}\label{sec:ETGs}
To select samples of passive, red, early-type galaxies, we followed two different approaches:
\begin{enumerate}
\item {\it Selection of red galaxies}\\
The first sample was selected by applying only a photometric classification (see also Pozzetti et al. 2009). Following Zucca et al. (2006),
we used the empirical set of 62 SEDs described in Ilbert et al. (2006). These SEDs were derived by interpolating between the four local observed spectra of Coleman et al. (1980) (from the old stellar population in M31 and M81 to Sbc, Scd, and Im SEDs) and two starburst SEDs from Kinney et al. (1996). These templates were linearly extrapolated to the ultraviolet ($\lambda < 2000$ \AA) and near-infrared wavelengths using the
GISSEL synthetic models (Bruzual \& Charlot 2003). Galaxy photometric types were evaluated by performing a best-fit modeling of the multi-band photometry using these templates.\\
Among these galaxies, we selected only those with a SED most closely fitted with photometric type = 1, which correspond to the 13 reddest E/Sa templates, obtaining a sample of 2098 galaxies (hereafter ``{\it red} galaxies").
\begin{figure}[t!]
\centering
\includegraphics[width=0.49\textwidth]{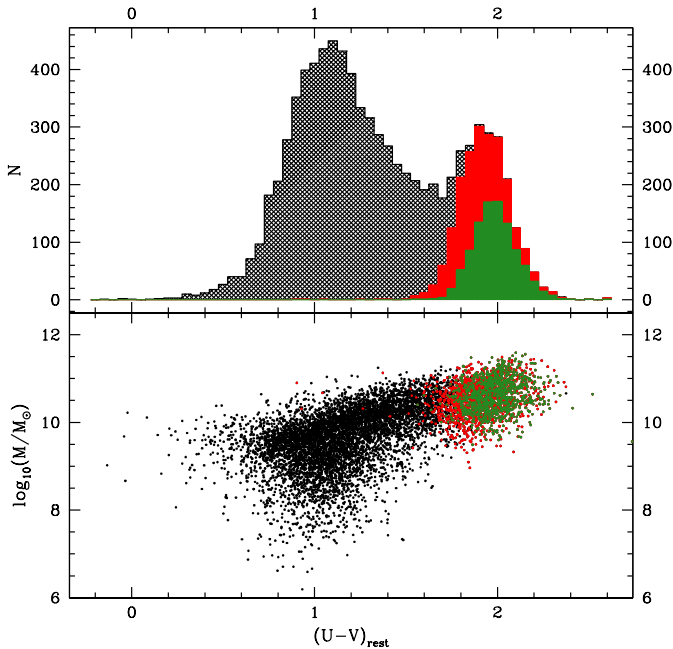}
\caption{Distribution of $(U-V)_{rest}$ color (upper panel) and color-mass relation (lower panel) for the parent zCOSMOS bright 10k galaxies sample (in black), {\it red} galaxies (in red), and {\it ETGs} (in green).}
\label{fig:fig11}
\end{figure}
\item {\it Spectrophotometric and morphological selection of ETGs}\\
 Every kind of selection of early-type galaxies is affected to some degree by contamination from outliers (e.g., see Franzetti et al. (2007) for a discussion of color-selected contamination). To test our results, we adopted a more restrictive criterion, based on a joint analysis of photometric, spectroscopic, and morphological data, to obtain a reliable selection of ETGs, biased as little as possible toward other types of galaxies.
Starting from the {\it red} sample, we chose to remove all galaxies with:
\begin{itemize}
\item {\it significantly strong emission lines}, i.e., measured \mbox{$EW([OII])>5$\AA} ($\sim17\%$ of the {\it red} sample) and $EW(H\alpha)>5$\AA ($\sim14\%$ of the {\it red} sample), to avoid selecting star-forming galaxies or galaxies that have experienced a recent episode of star formation. The threshold at 5 \AA was chosen on the basis of the results shown by Mignoli et al. (2009). They find a bimodality in the distribution of $EW([OII])$, and suggest that 5 \AA is a good choice of $EW([OII])$ value to separate starforming from passive galaxies, which are most commonly found below this value (see also Schiavon et al. 2006). In this way, we are confident about our selection of a clearly defined sample of passive galaxies.
\item {\it confirmed spiral morphology}: the morphologies of zCOSMOS galaxies were determined using two different types of classification, one performed by the Marseille group (Cassata et al. 2007, 2008; Tasca et al. 2009) and the other by the Zurich group (Scarlata et al. 2007). We decided to reject a galaxy when both groups classify it as spiral ($\sim20\%$ of the {\it red} sample) to be sure of a clear identification of the galaxy morphology.
\item {\it strong $24 \mu m$ emission}, i.e., based on an observed color $K-24 \mu m > -0.5$, which corresponds approximatively to a cut for a late-type template ($\sim5\%$ of the {\it red} sample).
\item {\it ``bluer" photometric types}, i.e., given the subdivision in photometric type of {\it red} galaxies from 1.01 (reddest) to 1.13 (bluest), we removed galaxies with photometric type $\geq1.05$ ($\sim31\%$ of the {\it red} sample) to discard among the remaining galaxies the bluer ones (see also Zucca et al. 2009).
\end{itemize}
In this way, we obtained 981 galaxies (hereafter ``{\it ETGs}").
\end{enumerate}
Finally, we also analyzed in detail the properties of {\it red galaxies} and {\it ETGs}. From a study of the color-mass diagram, we found that our sample is well placed in the red sequence (see Fig. \ref{fig:fig11}, lower panel), and that the sample of {\it ETGs} has a mean color that is even redder (see Fig. \ref{fig:fig11}, upper panel). Considering the mass distribution, we found that the {\it ETG} selection tends to include the more massive of the {\it red} galaxies (see Fig. \ref{fig:fig11}). We are thus confident that our conservative selection of {\it ETGs} has selected the more massive, passive, redder, and spheroidal galaxies of the zCOSMOS 10k-bright sample.

\begin{figure}[t!]
\centering
\includegraphics[width=0.49\textwidth]{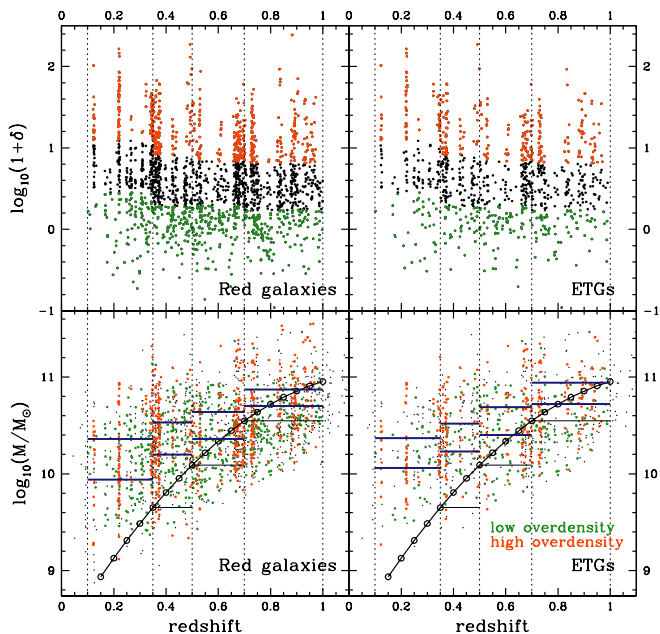}
\caption{Overdensity-redshift (upper panels) and mass-redshift (lower panels) distributions of the sample of {\it red} galaxies (left columns) and ETGs (right columns). In the overdensity-redshift distribution, the galaxies residing in the lowest overdensity quartile are highlighted in green, and in red the ones in the highest. In the mass-redshift distributions, we indicate by black dotted vertical lines the redshift division of the sample, by black horizontal lines the lower mass limit adopted in each redshift bin, and by blue horizontal lines the division into tertiles for each bin. The curve joining the open circles represents the mass completeness of the sample.}
\label{fig:fig3}
\end{figure}

\subsection {Overdensity definition}\label{sec:environment}

Determining an appropriate overdensity definition for the zCOSMOS 10k-bright sample is a difficult problem. For zCOSMOS, Kova\v{c} et al (2009) defined the density contrast to be $\delta=(\rho-\bar{\rho})/\bar{\rho}$, where $\rho$ is a function of the position of the galaxy (RA, dec, and z) and $\bar{\rho}$ is the mean value measured at the same redshift. The density field of the COSMOS field was reconstructed for different choices of filters (a fixed comoving aperture or an adaptive aperture with a fixed number of neighbors), tracers (flux-limited or volume-limited subsamples), and weights (stellar mass, luminosity, or no weight, i.e., considering only the number of galaxies); we refer to Kova\v{c} et al. (2009) for a comprehensive explanation.
Following the approach of Bolzonella et al. (2009), we used the 5th nearest neighbor volume-limited (5NN) estimator of the density field, because it has been shown that this represents a good compromise between the smallest accessible scales and the reliability of the overdensity values. We decided to use a number weighted estimator, and not a mass weighted one, because here we are trying to distinguish a dependence on mass, and introducing an estimator that itself depends on mass could bias our analysis. We find good agreement between the results obtained with this estimator and a luminosity weighted estimator. Hence the results discussed in this paper do not depend on the choice of the adopted estimator for the density field.
In Table \ref{tab:tab1}, we report the overdensity quartiles that were evaluated in 4 different redshift bins considering galaxies with $log_{10}(M/M_{\odot})>10.5$, and we refer to Sect. 3.4 of Bolzonella et al. (2009) for a detailed discussion. In the upper panels of Fig. \ref{fig:fig3}, we show the overdensity-redshift relation for the {\it red} and the {\it ETGs} sample, highlighting in green the lowest and in red the highest overdensity quartile.\\
Table \ref{tab:tab7} provides information about the median overdensity measured in the four environment quartiles and the distance scale measured in the radius within the 5NN. The change in the median overdensity from one quartile to the next is by a factor of $\sim2$.
Knobel at al. (2009) detected 102 groups with more than 5 observed members and 23 with more than 8 in the zCOSMOS 10k sample.  Finoguenov et al. (2007) analyzed the XMM-COSMOS survey, finding an X-ray luminosity for extended sources between $L_{0.1-2.4keV}=3\cdot10^{42} \mathrm{erg\,s^{-1}}$ and $5\cdot10^{43}\mathrm{erg\,s^{-1}}$. Peng et al. (2010) made a detailed comparison between zCOSMOS and SDSS, evaluating in the same way the overdensities in the two surveys, and finding that SDSS spans a wider range of overdensity, from $log_{10}(1+\delta)=-1$ to almost 3, while the overall zCOSMOS range is $-1<log_{10}(1+\delta)<2.3$. For ETGs, as shown from Fig. \ref{fig:fig3}, the range is a little smaller, with $-0.5<log_{10}(1+\delta)<2$.
We therefore note that, the zCOSMOS survey does not contain any single rich cluster, comparable to the Coma cluster in the local Universe for example (the highest mass in the Knobel et al. sample is $\approx10^{14}M_{\odot}$, while the mass of Coma cluster is $\approx10^{15}M_{\odot}$), but samples well the rich group environment (see also Iovino et al., 2009).

\begin{figure}[t!]
\centering
\includegraphics[width=0.49\textwidth]{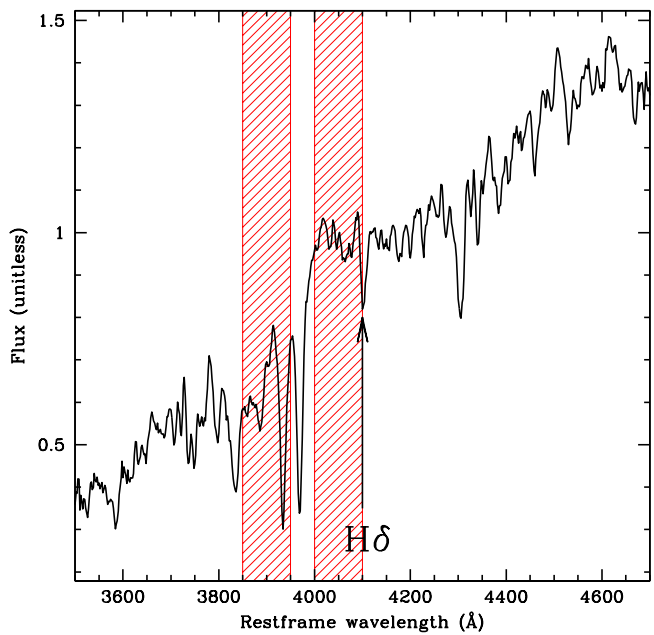}
\caption{Stacked spectrum of the 607 {\it ETGs} in the range $0.45<z<1$. In red, we plot the regions of the red and blue continuum used in the computation of D4000.}
\label{fig:fig16}
\end{figure}

\begin{table}[b!]
\begin{center}
\begin{tabular}{ccc}
\hline \hline
$\Delta z$ & $1^{st}$ quartile & $2^{nd}$ quartile\\
\hline
$0.1<z<0.35$ & $x\leq2.82$ & $2.82<x\leq5.36$\\
$0.35\leq z<0.5$ & $x\leq1.94$ & $1.94<x\leq3.57$\\
$0.5\leq z<0.7$ & $x\leq1.95$ & $1.95<x\leq3.52$\\
$z\geq0.7$ & $x\leq1.72$ & $1.72<x\leq3.39$\\
\hline
\end{tabular}
\begin{tabular}{ccc}
\hline
$\Delta z$ & $3^{rd}$ quartile & $4^{th}$ quartile\\
\hline
$0.1<z<0.35$ & $5.36<x\leq12.21$ & $x>12.21$\\
$0.35\leq z<0.5$ & $3.57<x\leq8.25$ & $x>8.25$\\
$0.5\leq z<0.7$ & $3.52<x\leq6.84$ & $x>6.84$\\
$z\geq0.7$ & $3.39<x\leq6.49$ & $x>6.49$\\
\hline \hline
\end{tabular}
\caption{Definition of quartiles for the overdensities computed with the 5NN estimator. The variable $x$ represents the overdensity $(1+\delta)$.}
\label{tab:tab1}
\end{center}
\end{table}
\begin{table}[b!]
\begin{center}
\begin{tabular}{ccc}
\hline \hline
quartile & median overdensity & median distance\\
\hline
$1^{st}$ & $1.23\pm0.03$ & $3.03\pm0.06$\\
$2^{nd}$ & $2.81\pm0.04$ & $1.83\pm0.03$\\
$3^{rd}$ & $5.29\pm0.07$ & $1.31\pm0.02$\\
$4^{th}$ & $14.43\pm0.38$ & $0.84\pm0.02$\\
\hline \hline
\end{tabular}
\caption{Median overdensities and distances (in comoving Mpc/h) with their errors for the four environment quartiles averaged throughout the entire redshift range.}
\label{tab:tab7}
\end{center}
\end{table}

%-------------------------------------------------------------------------------

\section{The analysis}\label{sec:analysis}

\subsection {Mass and color estimates}\label{sec:massbin}
The masses of {\it red} galaxies and {\it ETGs} were estimated by performing a best fit to the multicolor spectral energy distribution, using the observed magnitudes in 10 photometric bands from $u^*$ to $4.5\mu m$. Since we analyze a sample of red and passive galaxies, we decided to use different population synthesis models with short SFHs and low reddening; thus from Charlot \& Bruzual (2007) stellar population models (CB07, private communication), we adopted models with smoothly exponentially decreasing SFHs with $\tau<1$ Gyr, solar metallicity, and $0<A_{V}<1$.
To search for any mass dependence of the color distribution, we divided our sample into bins of stellar masses (Fig. \ref{fig:fig3} shows the different divisions adopted). Bolzonella et al. (2009) found that the mass function of ETGs does not vary significantly in the high and low-density environments probed by the zCOSMOS survey. The first step we therefore took was to simply divide the sample into equally populated mass tertiles over the entire redshift range for all the overdensity bins considered (each bin contains approximatively 200 galaxies). The lowest tertile corresponds to the mass range $log_{10}(M/M_{\odot})< 10.25$, the intermediate tertile to $10.25<log_{10}(M/M_{\odot})<10.6$, and the upper tertile to $log_{10}(M/M_{\odot})>10.6$. The choice of a division into mass tertiles instead of quartiles (as performed for the analysis of environment) is justified by our wish to keep the statistical accuracy high and the shot noise low.\\ Owing to the magnitude selection ($I_{AB}<22.5$), our sample is affected by mass incompleteness that increases with redshift (see Fig. \ref{fig:fig3}). To correct for the mass incompleteness effect, we applied the classical non-parametric $1/V_{max}$ formalism (Schmidt 1968). To analyze the redshift evolution, we also studied our samples divided into four redshift bins (identical to those adopted in the overdensity quartiles definition), evaluating the tertiles above the mass completeness limit in each redshift bin, as shown in the bottom part of Fig. \ref{fig:fig3}. In this case, every bin contains approximately 40 galaxies. This procedure yielded a lower statistical accuracy with the advantage of following in detail the redshift evolution.\\
Absolute magnitudes and colors were evaluated using the ALF software following the method described in Zucca et al. (2009); we analyzed the $(U-V)$ (Buser's filters, Buser 1978) rest-frame color.

\begin{figure*}[t!]
\centering
\includegraphics[angle=0,width=0.95\textwidth]{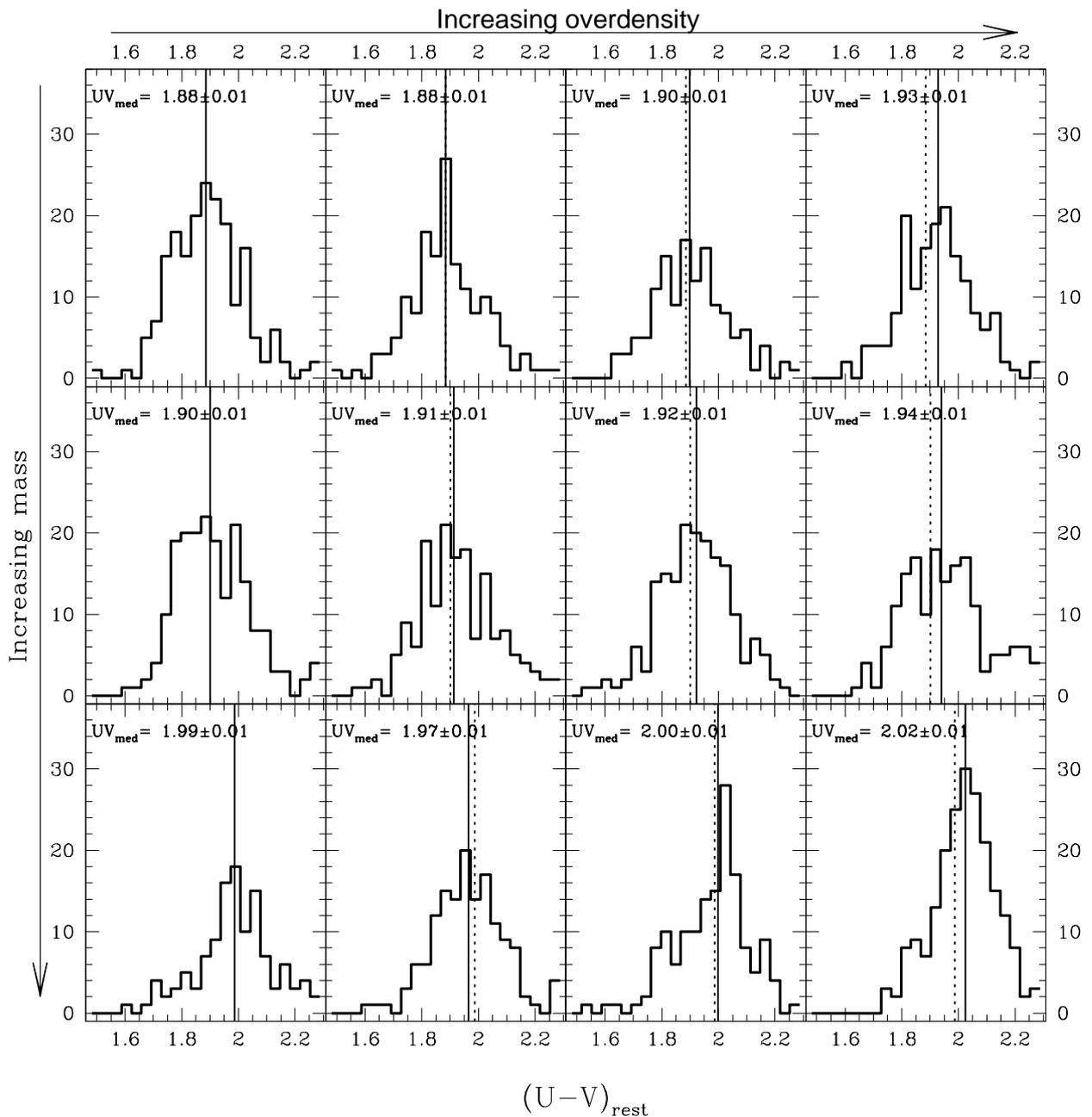}
\caption{$(U-V)$ rest-frame color distribution for {\it red} galaxies as a function of increasing overdensities (horizontal panels) and increasing mass (vertical panels). The mass was divided into three ranges, $log_{10}(M/M_{\odot})<10.25$, $10.25\le log_{10}(M/M_{\odot})<10.6$, and $log_{10}(M/M_{\odot})\ge10.6$ (panels from top to bottom), while the quartiles of environment are shown in Tables \ref{tab:tab1} and \ref{tab:tab7} (panels from left to right). In each panel, the solid line represents the median of the distribution, while the dashed line is the median value corresponding to the color distribution in the same mass range but for the lowest environment, to make it easier to follow the environment dependence. In each panel, we also report the value of the median $(U-V)_{rest}$ with its error.}
\label{fig:fig10}
\end{figure*}

\subsection {D4000 and $EW_{0}(H\delta)$ definitions} \label{sec:d4000}
Combinations of spectroscopic line strengths measurable in galaxy spectra have been identified to be direct tracers of galaxy characteristic parameters, such as mass, age, and star-formation rate. In particular, the $4000$ \AA break (D4000) and the rest-frame equivalent width ($EW_{0}$) of the $H\delta$ Balmer line are indicators of galaxy integrated or recent star formation, respectively. These same indicators were used in the zCOSMOS sample to study the activity of 24 $\mu m$ sources and their close neighbors (Caputi et al. 2009a, Caputi et al. 2009b) and to select post-starburst galaxies (Vergani et al. 2009).
To easily study the spectral properties of the zCOSMOS galaxies, we used of fully automatic spectral measurement code (PlateFit, Lamareille et al. 2006), which measures all the spectral features and equivalent widths of the prominent lines present in zCOSMOS spectra, given its spectral range. This code fits the stellar continuum and absorption lines making use of the STELIB library (Le Borgne et al. 2003) of stellar spectra and the GALAXEV (Bruzual \& Charlot 2003) stellar population synthesis models (see Lamareille et al. (2006) for further details). To investigate the influence of mass and environment on the spectral features, we decided to average the values of D4000 and $EW_{0}(H\delta)$ evaluated for single spectra in bins of redshift, mass, and environment. Because of the wavelength coverage of the zCOSMOS spectra, the D4000 index is available only for $0.45\leq z\leq1$, which corresponds approximately to a time interval of 4 Gyrs. To follow the redshift evolution, we divided this redshift range into four bins ($0.45\le z<0.6$, $0.6\le z<0.7$, $0.7\le z<0.8$, and $0.8\le z\le1$). We divided each redshift bin yet further using the four environment quartiles defined as before, to ensure a coherent analysis, and three equally populated mass tertiles, with the lowest one corresponding to  $log_{10}(M/M_{\odot})<10.4$, the intermediate to $10.4<log_{10}(M/M_{\odot})<10.7$, and the highest one to $log_{10}(M/M_{\odot})>10.7$. As a starting catalog, we used the {\it ETG} sample, because it is less contaminated by outliers. Figure \ref{fig:fig16} shows the average stacked spectrum of the 607 {\it ETGs} in the range $0.45<z<1$, the ranges of the blue and red bands used to evaluate D4000 and $H\delta$ indices. On average each bin contains 10 galaxies.
We adopted the D4000 definition given by Balogh et al. (1999), where the average flux $F_{\nu}$ in the red and blue wavelength bands are measured in the ranges $4000-4100$ \AA and $3850-3950$ \AA (D4000 (narrow)):
\begin{equation}
D4000=\frac{F_{red}}{F_{blue}}=\frac{(\lambda_{2}^{blue}-\lambda_{1}^{blue})\int_{\lambda_{1}^{red}}^{\lambda_{2}^{red}}F_{\nu}d\lambda}{(\lambda_{2}^{red}-\lambda_{1}^{red})\int_{\lambda_{1}^{blue}}^{\lambda_{2}^{blue}}F_{\nu}d\lambda}.\nonumber
\end{equation}

%-------------------------------------------------------------------------------

\section{Results}\label{sec:res}

\subsection{Color distribution of early-type galaxies}
The $(U-V)_{rest}$ color distributions for the total sample of {\it red} galaxies are shown in Fig. \ref{fig:fig10}. We evaluated the color distributions and their medians in four quartiles of environment and three mass bins, as described in Sects. \ref{sec:environment} and \ref{sec:massbin}, and plotted them in order of both increasing mass (from top to bottom) and increasing density environment (from left to right). In each panel, we also indicated by a dashed line the median value corresponding to the color distribution in the same mass range but for the lowest density environment, in order to make it easier to follow the environment dependence. It is possible to see that, although the dispersion of the color distributions is almost the same in different mass and environment bins, there is evolution in the median color with mass for fixed environment, the medians shifting towards a redder color with increasing mass. On the other hand, if we focus on environment we see that its effect at fixed mass is not strong, there being only a small shift in the median value mostly for the highest density quartile. In the following sections, we explore these dependences quantitatively, studying also the effect of redshift evolution.

\subsection{Color-environment relation}
\begin{figure}[t!]
\centering
\includegraphics[width=0.49\textwidth]{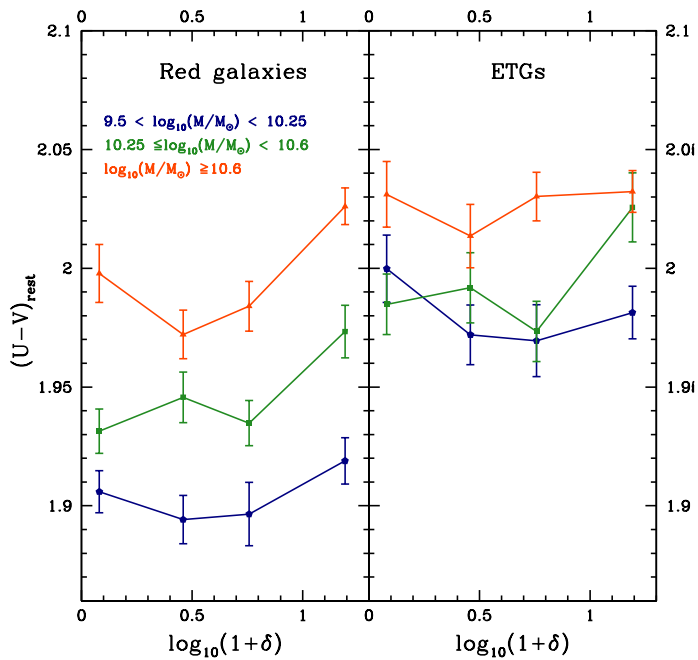}
\caption{Mean color of {\it red} galaxies and {\it ETGs} as a function of environment in different mass bins. In the x-axis, we plot the quartile of overdensity, from the lowest (1) to the highest (4) value as shown in Table \ref{tab:tab1}.}
\label{fig:fig4}
\end{figure}

We evaluated the mean color of {\it red} galaxies and {\it ETGs} in the four quartiles of environment defined in Table \ref{tab:tab1}; the results are plotted in Fig. \ref{fig:fig4} for the three mass bins. As explained in Sect. \ref{sec:massbin}, we applied a $1/V_{max}$ correction to take into account the incompleteness in mass, evaluating the mean color by weighting each galaxy by its $V/V_{max}$, so the values reported in Figs. \ref{fig:fig4} and \ref{fig:fig5} differ from to the values shown in Fig. \ref{fig:fig10}. It is evident that within a fixed mass bin the dependence of the mean color on the overdensity parameter for both of the samples is really weak, showing on average an offset of $\left<\Delta(U-V)_{rest}\right>=0.027\pm0.008$ mag redder in the highest density environments than in the lowest density environment for {\it red} galaxies and $\left<\Delta(U-V)_{rest}\right>=0.01\pm0.01$ mag for {\it ETGs}.
\begin{table}[b!]
\begin{center}
\begin{tabular}{ccc}
\hline \hline
 & {\it Red sample} & {\it ETGs}\\
\hline
low mass & $0.012\pm0.006$ & $-0.011\pm0.008$\\
medium mass & $0.031\pm0.006$ & $0.028\pm0.008$\\
high mass & $0.039\pm0.006$ & $0.007\pm0.007$\\
\hline \hline
\end{tabular}
\caption{Slopes $S_{\delta}$ of the color-environment relation \mbox{$(U-V)_{rest}\propto S_{\delta}\cdot log_{10}(1+\delta)$} in different mass bins for {\it red} galaxies and {\it ETGs}.}
 \label{tab:tab2}
\end{center}
\end{table}
We evaluated the formal slopes $S_{\delta}$ of this relation \mbox{$(U-V)_{rest}\propto S_{\delta}\cdot log_{10}(1+\delta)$} within our overdensity range, which spans the values $0.1\lesssim1+\delta\lesssim200$, and the results are shown in Table \ref{tab:tab2}. All the slopes evaluated for the {\it red galaxy} sample are quite shallow, always being below a value of $S_{\delta}\approx0.04$, with a hint of steepening with increasing mass. In the {\it ETG} sample, which is 50\% smaller, the slopes are rather consistent with being flat.\\

\begin{table}[b!]
\begin{center}
\begin{tabular}{ccc}
\hline \hline
 & {\it Red sample} & {\it ETGs}\\
\hline
low env & $0.11\pm0.01$ & $0.04\pm0.01$\\
med low env & $0.11\pm0.01$ & $0.06\pm0.02$\\
med high env & $0.12\pm0.01$ & $0.09\pm0.01$\\
high env & $0.14\pm0.01$ & $0.07\pm0.01$\\
\hline
all env & $0.126\pm0.005$ & $0.066\pm0.007$\\
\hline \hline
\end{tabular}
\caption{Slopes $S_{M}$ of the color-mass relation \mbox{$(U-V)_{rest}\propto S_{M}\cdot log_{10}(M/M_{\odot})$} in different environment bins for {\it red} galaxies and {\it ETGs}.} \label{tab:tab4}
\end{center}
\end{table}

\subsection{Color-mass relation}\label{sec:colmass}
\begin{figure}[t!]
\centering
\includegraphics[width=0.49\textwidth]{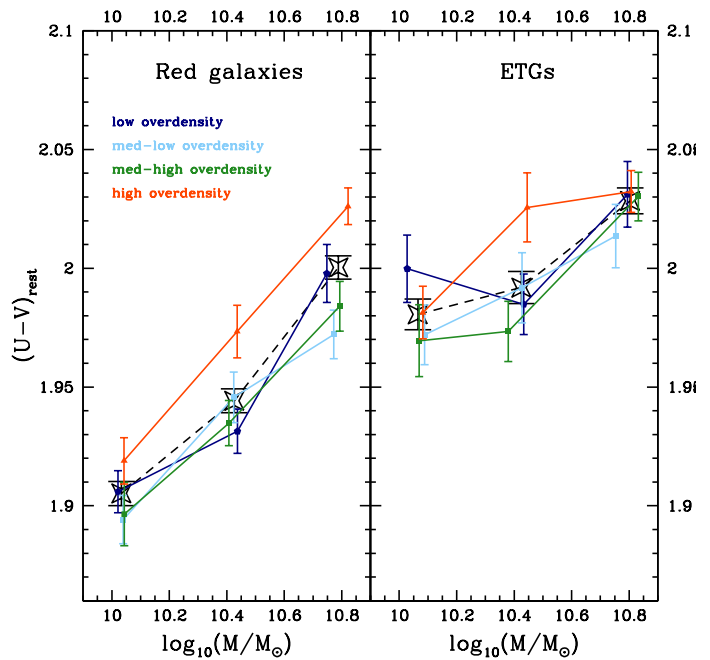}
\caption{Mean color of {\it red} galaxies and {\it ETGs} as a function of mass in the four quartiles of environment. Black stars connected with dashed lines show the mean colors for the entire {\it red galaxy} and {\it ETG} sample, not divided into quartiles of environment.}
\label{fig:fig5}
\end{figure}

\begin{figure}[t!]
\centering
\includegraphics[width=0.49\textwidth]{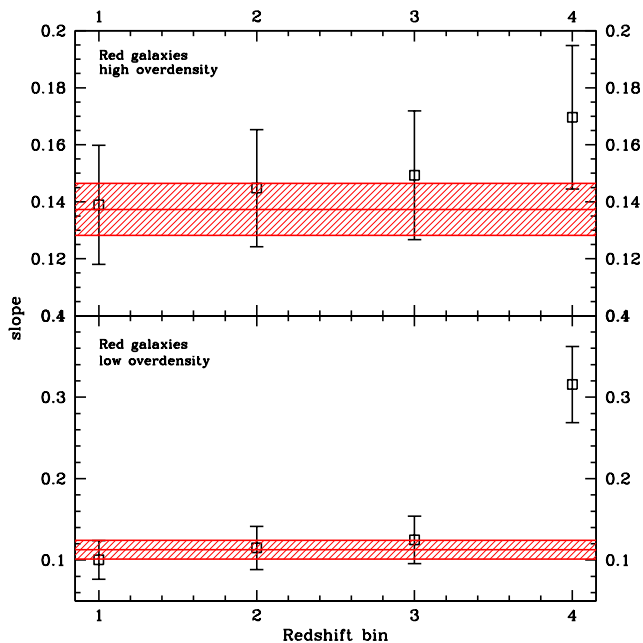}
\caption{Slopes of the color-mass relation evaluated in the four redshift bins and different environments. The red lines represent the slope of the color-mass relation evaluated for the entire redshift sample in the corresponding overdensity bin.}
\label{fig:fig6}
\end{figure}

To study the color-mass relation of early-type galaxies, we computed the mean of the color distribution as a function of mass in the four quartiles of environment. Figure \ref{fig:fig5} presents our results. From these plots, it is possible to see that for the sample of {\it red} galaxies there is a mean reddening of $\left<\Delta(U-V)_{rest}\right>=0.093\pm0.007$ mag between the lowest ($9.5<log_{10}(M/M_{\odot})<10.25$) and highest masses ($log_{10}(M/M_{\odot})>10.6$), while for the {\it ETGs} the difference is smaller ($\left<\Delta(U-V)_{rest}\right>=0.047\pm0.009$). Furthermore, in Fig. \ref{fig:fig5} the mean of the color distribution shown in black, evaluated for the whole sample, shows that the slopes of the relation are similar in different environments.\\
The slopes $S_{M}$ of the relation \mbox{$(U-V)_{rest}\propto S_{M}\cdot log_{10}(M/M_{\odot})$} found in the mass range of our sample, with mean values in the range $10<log_{10}(M/M_{\odot})<10.8$, are reported in Table \ref{tab:tab4}. The values of the slopes confirm a rather strong dependence of the color on mass, presenting also a small steepening with increasing environment. The mean value of the slope is $\left<S_{M}\right>=0.126\pm0.005$ for the {\it red galaxy} sample. The slopes found for the {\it ETG} sample are a little shallower ($\left<S_{M}\right>=0.066\pm0.007$), probably because the stricter color selection applied to obtain the {\it ETGs} sample reduces the range of variation in the $(U-V)_{rest}$ color.
In the present study, we didn't correct our colors for redshift evolution. To verify that our results are not biased by this effect, we studied the {\it red} galaxies, since they are the most numerous sample, dividing them into four redshift bins, $0.1<z<0.35$, $0.35<z<0.5$, $0.5<z<0.7$, and $0.7<z<1$; in each bin, we then evaluated equally populated tertiles of mass, and studied the slope of the resulting color-mass relation in the respective redshift bin. The result is shown in Fig.\ref{fig:fig6}: the open squares represent the slope of the relation in each redshift range with its error, while the red shaded area represents the global slope found in the same environment when we do not divide the sample into redshift bins. In seven out of eight bins, the slope is consistent, at $1\sigma$ level, with the one evaluated over the entire range of redshift, showing only a hint of steepening in the last redshift bin (see also Table \ref{tab:tab3}).
\begin{table}[b!]
\begin{center}
\begin{tabular}{ccc}
\hline \hline
 & low env & high env\\
\hline
$z<0.35$&$0.10\pm0.02$&$0.14\pm0.02$\\
$0.35<z<0.5$&$0.11\pm0.03$&$0.14\pm0.02$\\
$0.5<z<0.7$&$0.12\pm0.03$&$0.15\pm0.02$\\
$z>0.7$&$0.32\pm0.05$&$0.17\pm0.03$\\
\hline \hline
\end{tabular}
\caption{Slopes $S_{M}$ of the color-mass relation for {\it red} galaxies in high and low environment and in given redshift bin.} \label{tab:tab3}
\end{center}
\end{table}
These results confirm those of Balogh et al. (2004) for the nearby Universe. They analyzed the SDSS-DR1 survey, finding a strong dependence of $(u-r)_{rest}$ color on luminosity with a difference of $\approx 0.25$ mag between the ranges $-23<M_{r}<-22$ and $-19<M_{r}<-18$, with a corresponding slope $\approx0.15$. Their dependence on environment is much smaller, $\sim0.03-0.06$ for a projected local density in the range $0.1-10$, which corresponds to a slope $\sim0.015-0.03$ that closely agrees with the results of our analysis.
Cooper et al. (2010) analyzed the $(U-B)_{rest}$ color searching for a dependence on environment for galaxies from the DEEP2 sample, in the mass range $10.6<log_{10}(M/M_{\odot}<11.1$) and the redshift range $0.75<z<0.95$. They found that galaxies residing in high density environments have slightly redder colors than galaxies in low density environments; however, here we demonstrate that the dependence on the mass is much stronger.

\subsection{Spectroscopic indices}\label{sec:spectral}
We measured the $D4000$ and the $EW_{0}(H\delta)$ for the {\it ETG} sample as described in Sect. \ref{sec:d4000} by analyzing the mean of the values of the spectral measurements evaluated in each galaxy spectrum, and averaging in fixed redshift, mass, and environment bins, as explained in Sect. \ref{sec:d4000}.
\begin{figure}[t!]
\centering
\subfigure{\includegraphics[angle=-90,width=0.49\textwidth]{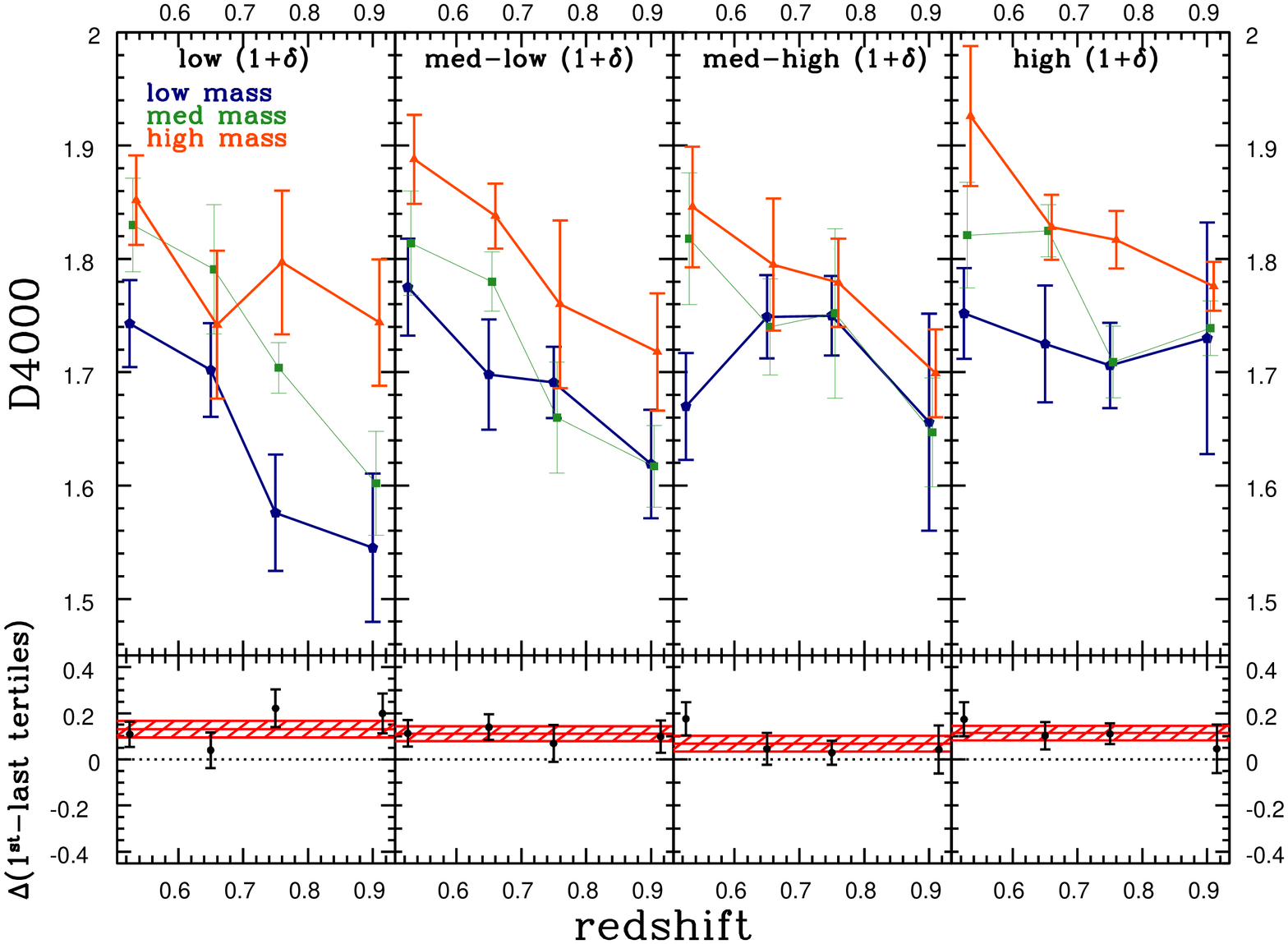}}
\subfigure{\includegraphics[angle=-90,width=0.49\textwidth]{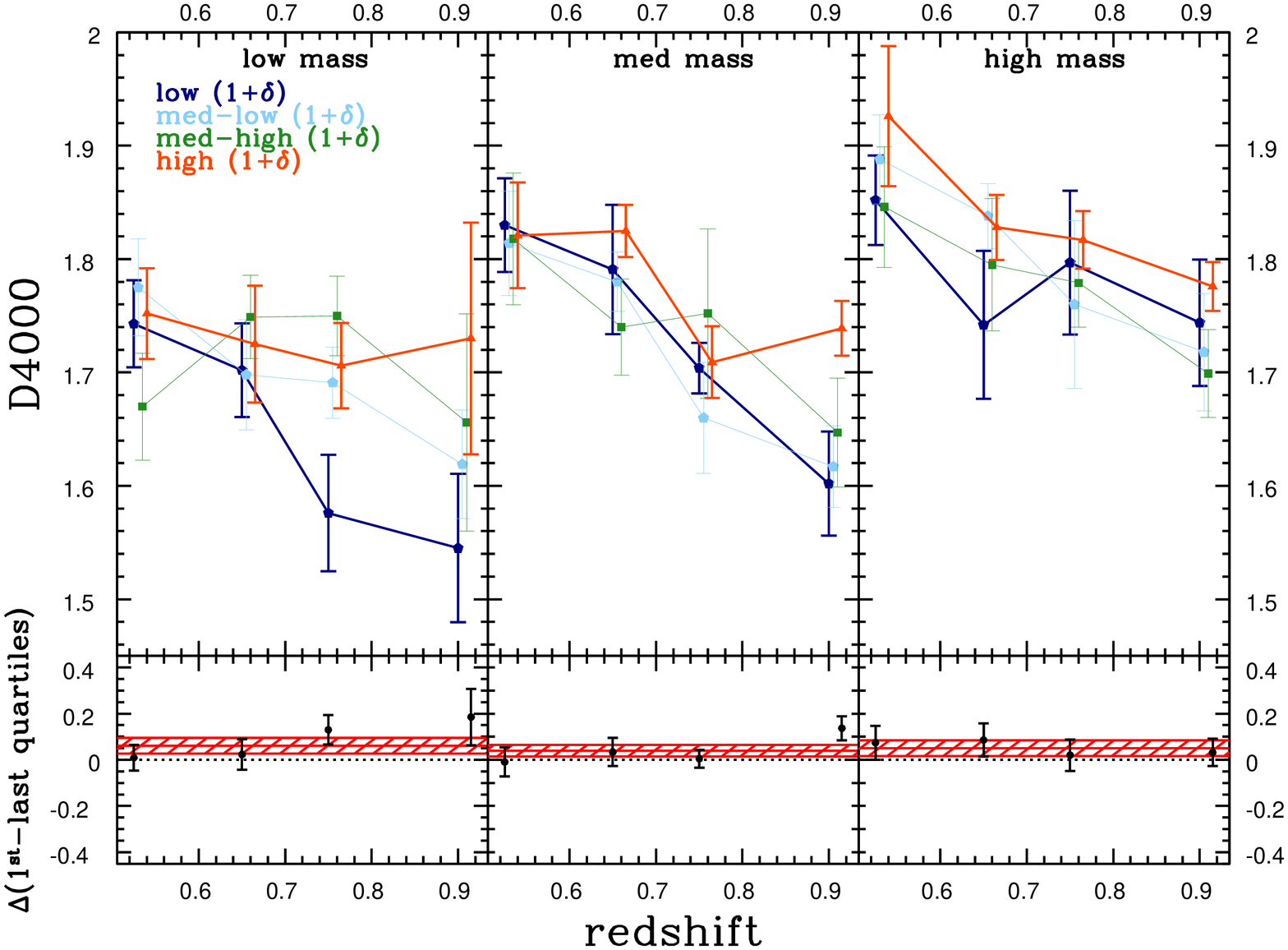}}
\caption{$D4000$ narrow evaluated for different mass bins in fixed environment quartiles (upper panels) and for different overdensity quartiles in fixed mass bins (lower panels). At the bottom of each panel, we also plot the differences between D4000 in extreme mass/overdensity ranges, the red shaded area representing the weighted mean of these differences averaged over the redshift range (see Table \ref{tab:tab5}). }
\label{fig:fig12}
\end{figure}

\begin{figure}[h!]
\centering
\subfigure{\includegraphics[angle=-90,width=0.49\textwidth]{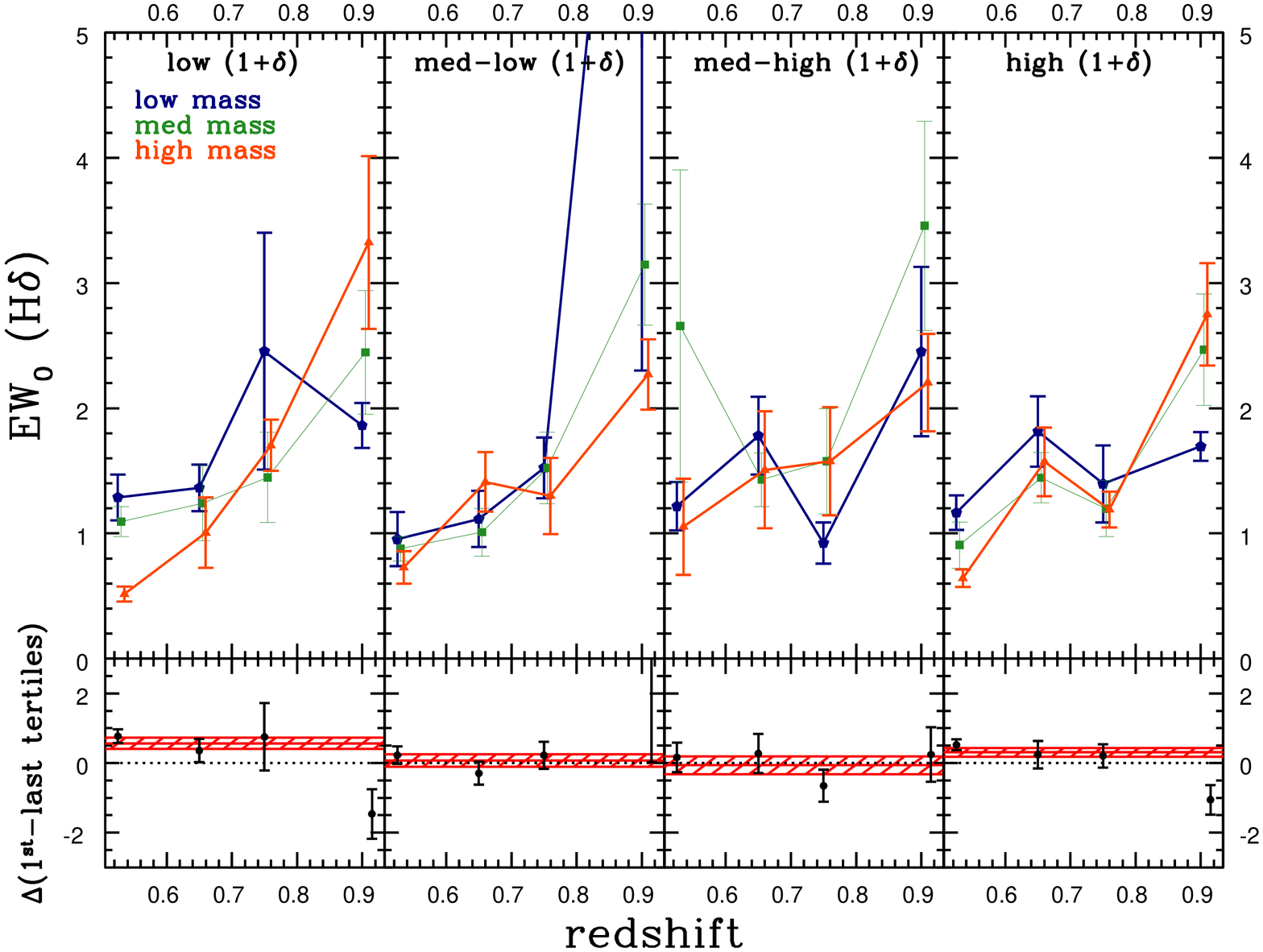}}
\subfigure{\includegraphics[angle=-90,width=0.49\textwidth]{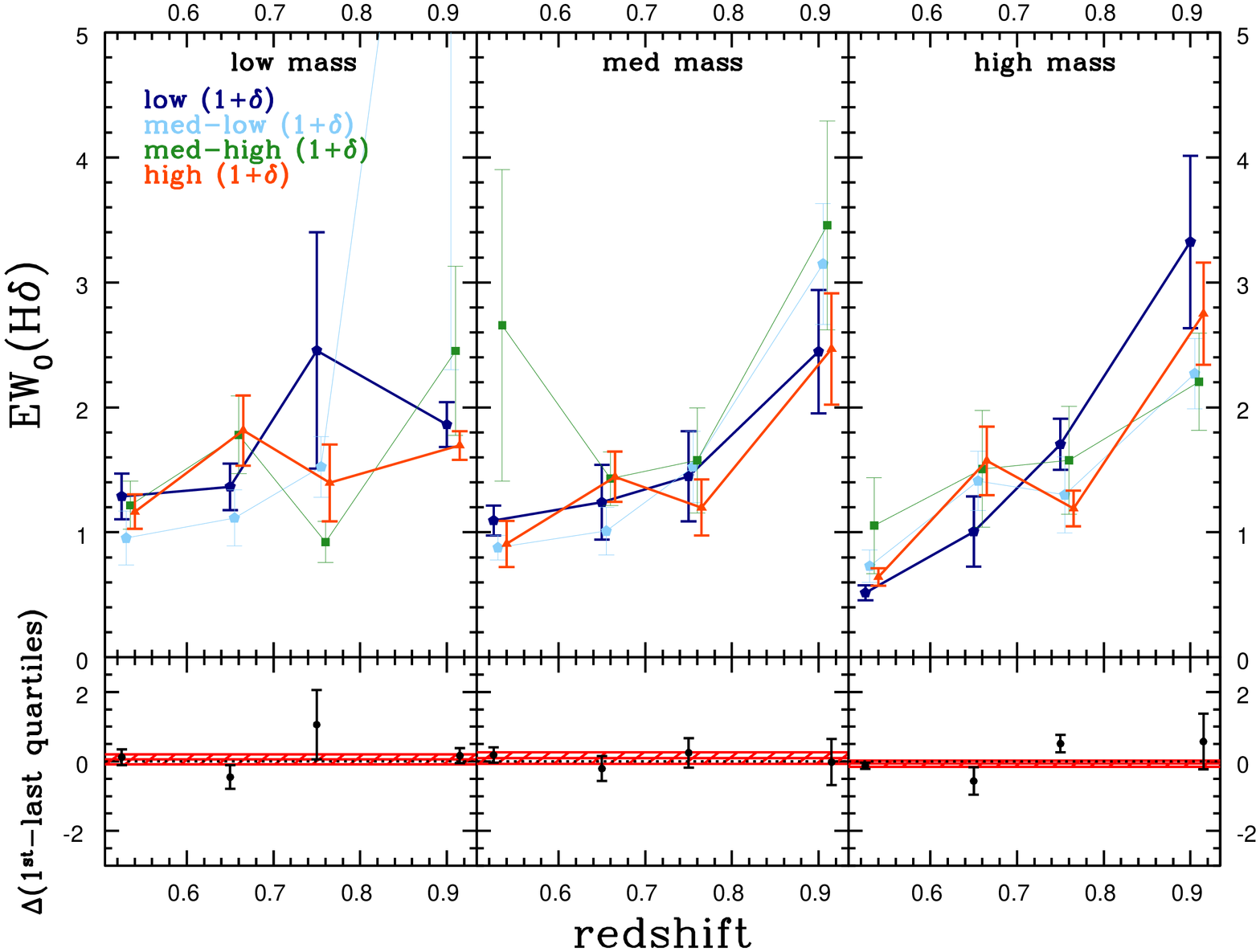}}
\caption{$EW_{0}(H\delta)$ evaluated for different mass bins in fixed environment quartiles (upper panels) and for different environment quartiles in fixed mass bins (lower panels). At the bottom of each panel, we also plot the differences between $EW_{0}(H\delta)$ in extreme mass/overdensity ranges, the red shaded area being the weighted mean of these differences averaged over the redshift range (see Table \ref{tab:tab6}).}
\label{fig:fig14}
\end{figure}
The results of the analysis of $D4000$ as a function of redshift are shown in Fig. \ref{fig:fig12}. In the upper panel, we keep the environment quartiles fixed and study the dependence on mass, while in the second panel, we study instead the dependence on environment at fixed mass. in all the mass and overdensity bins, we detect a redshift evolution in $D4000$, with the galaxies at low redshift always having a higher break value than at high redshift. This tells us that higher redshift galaxy populations are statistically younger and/or have a higher metallicity. In searching for any mass dependence, we find that massive galaxies exhibit a stronger D4000 break in all environments. The effect of environment is smaller, showing a trend in which galaxies residing in high overdensity regions have, for all mass bins, a greater D4000 value than galaxies in the lowest environment quartile.
\begin{table}[b!]
\begin{center}
\begin{tabular}{ccccc}
\hline \hline
 & low env & med-l env & med-h env & high env\\
\hline
$\left<\Delta D4000 \right>_{mass}$ & 0.13 & 0.11 & 0.07 & 0.11\\
$\#\sigma$&3.8&3.5&2&3.7\\
\hline
\end{tabular}
\begin{tabular}{cccc}
 \hline
 & low mass & med mass & high mass\\
\hline
$\left<\Delta D4000 \right>_{env}$ & 0.06 & 0.04 & 0.05\\
\#$\sigma$&1.8&1.5&1.5\\
\hline \hline
\end{tabular}
\caption{Differences between $D4000$ for galaxies of high and low mass in a given environment (upper table) and galaxies in high and low density environment quartiles at fixed mass (lower table), and  their respective significances.} \label{tab:tab5}
\end{center}
\end{table}
To study the influence of the mass and overdensity, we evaluated the differences of D4000 and $EW_{0}(H\delta)$ between the extreme quantiles of mass and environment and their significance. The black points in the lower panels of each figure in Fig. \ref{fig:fig12} show the difference in D4000 between high and low mass or environment regimes. To derive a mean value for each mass and environment regime, we averaged the differences found in each redshift bin using their errors as weights; the red shaded area in Fig. \ref{fig:fig12} represents the mean difference integrated as a function of redshift. We defined the significance to be the distance of this integrated difference of D4000 between extreme quartiles of mass and environments in units of $\sigma$, where $\sigma$ is the error in the integrated difference. \mbox{Table \ref{tab:tab5}} shows that the results are in agreement with those found in our color analysis: the dependence on mass is strong, with a mean $\left<\Delta D4000\right>=0.11\pm0.02$ in a mass range $10.2\lesssim log_{10}(M/M_{\odot})\lesssim10.8$ and a significance over $3\sigma$ level (except for one overdensity bin), while the dependence on environment is weaker, with a mean $\left<\Delta D4000\right>=0.05\pm0.02$ and much lower significances.\\
\begin{table}[b!]
\begin{center}
\begin{tabular}{ccccc}
\hline \hline
 & low env & med-low env & med-high env & high env\\
\hline
$\left<\Delta H\delta \right>_{mass}$ & 0.56 & 0.07 & -0.06 & 0.31\\
$\#\sigma$&3.5&0.4&0.2&2.4\\
\hline
\end{tabular}
\begin{tabular}{cccc}
 \hline
 & low mass & med mass & high mass\\
\hline
$\left<\Delta H\delta \right>_{env}$& 0.06 & 0.10 & -0.06\\
\#$\sigma$&0.4&0.6&0.8\\
\hline \hline
\end{tabular}
\caption{Differences (in \AA) between $EW_{0}(H\delta)$ for galaxies of high and low mass in a given environment (upper table) and galaxies in high and low environment quartiles at fixed mass (lower table), and the relative significances.} \label{tab:tab6}
\end{center}
\end{table}
The mass incompleteness of our spectral analysis is not a problem. Since we decided to divide our sample with a fixed mass cut as a function of redshift, the effect is a slight shift in the median value of mass at all redshifts in each mass bin, so that, for example, the median mass of low-mass galaxies change from $log_{10}(M/M_{\odot})=10.1$ to $log_{10}(M/M_{\odot})=10.25$ with increasing redshift. This shift in mass would have to be taken into account if one were trying to precisely establish the redshift evolution of D4000 at fixed mass. However, in our analysis we are only interested in identifying the main parameter relating mass to environment, so this shift should not affect our results. Moreover, even by correcting for it considering the relative lack of massive galaxies at low redshift and of less massive ones at high redshift, would at most increase the range of mass explored in the different redshift bins.\\
A similar consideration has to be made for the metallicity. As pointed out before, the D4000 index depends on both the age and metallicity of a galaxy. Over the mass range probed here, we expect top find a metallicity range of $-0.1<log_{10}(Z/Z_{\odot})< 0.1$ (Gallazzi et al 2006, assuming modest metallicity evolution to $z\sim0.5$); this corresponds to a variation in D4000 of approximatively $\pm0.1$ for the models of Bruzual \& Charlot (2003) at very old ages and $\pm0.06$ at ages of 5 Gyr, which is similar to the change we detect. Hence, on the basis of these data, we are unable to distinguish whether this change is due to a variation in either age or metallicity or both. Our sole interest here is not to differentiate the effect of mass from those of metallicity in producing the trends we find, but to determine whether the difference in D4000 is, at each redshift, far more significant when we divide our sample into mass bins, or into environment bins. For the present analysis, our inability to differentiate between age and metallicity is therefore not a problem.\\
The results of the $EW_{0}(H\delta)$ analysis are a bit noisier, as shown in Fig. \ref{fig:fig14} and Table \ref{tab:tab6}. However, the trends found agree with the results of both the colors and D4000 analysis: the differences with mass ($\langle\Delta EW(H\delta)\rangle=0.28\pm0.08$\AA) have on average higher significance than the difference with environment ($>2\sigma$ in two bins over four), being these second ones not significant in every bin considered ($\langle\Delta EW(H\delta)\rangle=0.01\pm0.06$\AA).\\
In contrast to D4000, $EW_{0}(H\delta)$ increases with redshift and decreases with increasing mass and environmental density.
Moreover, very massive galaxies or galaxies residing in high density environments have an almost constant $EW_{0}(H\delta)$ throughout the entire
redshift range, of a value that indicates in these galaxies no recent episode of star formation has occurred within a timescale of $\approx5$ Gyrs.

\subsection{Signatures of downsizing}
\begin{figure}[t!]
\centering
\includegraphics[angle=0,width=0.49\textwidth]{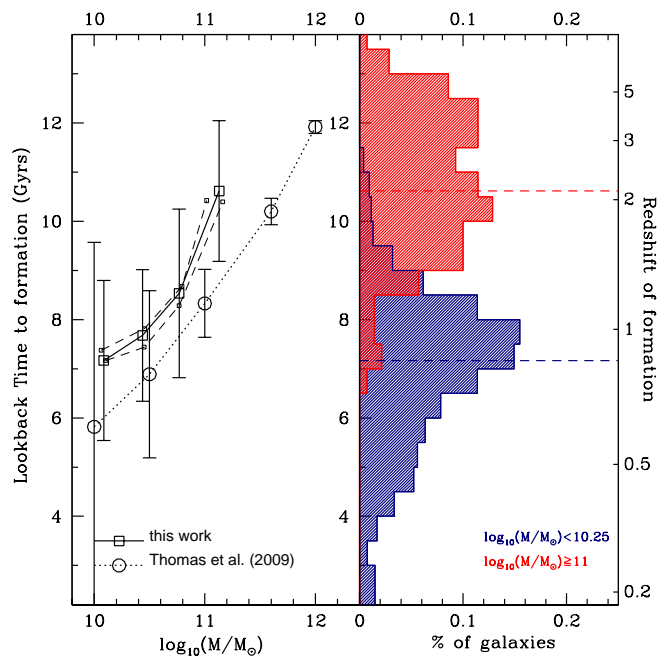}
\caption{Stellar mass as a function of lookback time to formation (left panel) and mass histograms of extreme tertiles for {\it ETGs}. Squared symbols represent our data, where the errorbars plotted are the dispersions in the measurements. Circle points are data taken from the analysis of SDSS-DR4 by Thomas et al. (2009), showing the lookback time at which the 50\% of the stellar mass has formed; the errorbars plotted are the differences between the lookback time of formation of 50\% and of 80\% of the stellar mass. The dashed lines represent the mass-lookback time relation computed in the extreme quartiles of overdensity. To compare data here we adopted $\Omega_{m}=0.24$, $\Omega_{\Lambda}=0.76$, and $H_{0}=73\,\mathrm{km\,s^{-1}Mpc^{-1}}$.}
\label{fig:fig18}
\end{figure}
The analysis of ETGs shows clear indications of age-downsizing, i.e. massive galaxies contain an older stellar population. We evaluated the formation redshift of {\it ETGs} by performing a best fit to the SED of these galaxies using a library of models with exponentially decreasing star-formation histories, with $\tau<1$ Gyrs, and we found high age/$\tau$ values compatible with passive galaxies. To check the robustness of our results, we also considered different stellar population synthesis models, such as the models of Bruzual \& Charlot (2003, BC03), and the more recent ones of Maraston (2005) and Charlot \& Bruzual (2007), which include the contribution from TP-AGB stars (respectively M05 and CB07). In Fig. \ref{fig:fig18}, we show the lookback time to formation as a function of mass for {\it ETGs}, evaluated using CB07 models; both the histogram and the plot show that massive galaxies ($log_{10}(M/M_{\odot})\sim11$) formed their stars at $z\sim2$, while less massive ones ($log_{10}(M/M_{\odot})\sim10.25$) at $z\lesssim1$. The corresponding mean age difference between the extreme mass quartiles is $\approx3$ Gyrs, while the age difference between different environments is insignificant ($\lesssim0.2$ Gyrs). The results from different models differ slightly, but the general picture is clear and any differences are within the $1\sigma$ errors. In Fig. \ref{fig:fig18}, we also plot the relation between mass and lookback time for the two extreme overdensity quartiles (dashed lines), showing that galaxy formation age at fixed mass is not strongly dependent on  environment. Our results agree with Thomas et al. (2009), which was based on an analysis of SDSS-DR4 early-type galaxies. As in our analysis, they found that their scaling relations had no strong on environment, but provided support for a downsizing in formation age with mass. For a comparison, we overplot in Fig. \ref{fig:fig18} their relation, where the points represent the lookback time at which the 50\% of the stellar mass has formed and the errorbars are the differences between the lookback time of formation for 50\% and 80\% of the stellar mass.\\
An interesting hint of downsizing can be inferred from the analysis of D4000 as a function of redshift. As can be seen in Fig. \ref{fig:fig12}, more massive galaxies at each redshift have an higher D4000 value than to less massive galaxies. However, given the magnitude of the errors in D4000, and our inability to clearly distinguish the effect of metallicity on D4000, we conclude that this provides confirmation only about the analysis on the ages of ETGs.

%-------------------------------------------------------------------------------

\section{Summary and conclusions}\label{sec:concl}

We have studied the (U-V) colors and spectra of two subsamples of early-type galaxies extracted from the \mbox{zCOSMOS} spectroscopic survey in the redshift range $0.1<z<1$. We have created two samples to complete a more robust and accurate statistical analysis of {\it red galaxies} ($\sim2100$ galaxies), and performed a more reliable selection of elliptical, red passive galaxies in the {\it ETG} case ($\sim1000$ galaxies) by combining spectroscopic, morphological, and photometric properties.
We have analyzed both color and spectral features (i.e., D4000 and $EW_{0}(H\delta)$) to explore their dependence on mass and environment and provide some insights into the main drivers of galaxy evolution.\\
From the analysis of the $(U-V)_{rest}$ color, we have found:
\begin{itemize}
\item a dependence on mass ($\left<\Delta(U-V)_{rest}\right>=0.093\pm0.007$ mag in the mass range $10<log_{10}(M/M_{\odot})<10.8$);
\item a weaker dependence on the local density of the environment ($\left<\Delta(U-V)_{rest}\right>=0.027\pm0.008$ mag in the overdensity range $0.1\lesssim log_{10}(1+\delta)\lesssim1.2$);
\end{itemize}
The mean slope of the color-mass relation is $\left<S_{M}\right>=0.12\pm0.005$, with a hint of steepening as a function of increasing environment. The color-environment relation is much flatter, with a slope that is always smaller than $S_{\delta}\approx0.04$, confirming evidence that the dependence of galaxy properties on mass is much stronger than on environment.
From the analysis of both samples, we have obtained results that are in complete agreement. The study of our {\it ETG} sample, carefully selected to be less contaminated by outliers, enhances the reliability of our conclusions.\\
Many previous efforts have been made to differentiate between the effects of mass and environment in studying the colors of ETGs. Our results agree with Balogh et al. (2004), who found by analyzing the SDSS-DR1 survey, a strong dependence of ETG color on luminosity (measuring a difference of $\approx 0.25$ mag between the ranges $-23<M_{r}<-22$ and $-19<M_{r}<-18$, in the $(u-r)_{rest}$ color) and a much smaller dependence on environment ($\lesssim0.1$ for a projected local density in the range $0.1-10$). We effectively extend these results to $z\sim1$. For the color-environment relation, Cooper et al. (2010) found a similar trend by analyzing galaxies observed by the DEEP2 survey in the mass range $10.6<log_{10}(M/M_{\odot})<11.1$ and in the redshift range $0.75<z<0.95$. Our results agreement with their findings that at fixed mass the $(U-B)_{rest}$ color in high density environments is slightly redder than in low density environments.\\
From the spectral analysis, we have found:
\begin{itemize}
\item a dependence of spectroscopic index on mass ($\left<\Delta D4000\right>=0.11\pm0.02$ and $\Delta EW_{0}(H\delta)=0.28\pm0.08$\AA in the mass range $10.2<log_{10}(M/M_{\odot})<10.8$) when averaging over the entire redshift range;
\item a weaker dependence on the local density of the environment ($\left<\Delta D4000\right>=0.05\pm0.02$ and an insignificant difference for $EW_{0}(H\delta)$ in the environment range $0.1\lesssim log_{10}(1+\delta)\lesssim1.2$) when averaging over the entire redshift range.
\end{itemize}
Massive galaxies or galaxies residing in relatively high density environments have a stronger D4000 break and a small value of $EW_{0}(H\delta)$ that is almost constant with redshift, implying that no significant episodes of star formation have occurred in these galaxies since $z\sim1$, while for the low-mass galaxies the evolution remains ongoing at these redshifts. The different amplitude of the redshift evolution in colors and spectral features can be interpreted by comparison with stellar population synthesis models: old, red, passive galaxies (modeled with values of age/$\tau$ ratio greater than 5) are not expected to have evolved much in color since z=1, while the evolution in D4000 is a bit clearer.\\
Combining the results of the color and spectral analysis, we conclude from this work that the main driver of the the evolution of ETGs is mass, while environment plays a less important although non-negligible role. This result is also confirmed by the analysis of Cucciati et al. (2010); from their study of the $(U-B)_{rest}$ color of the zCOSMOS bright sample, they suggest a scenario in which the color depends primarily on mass, but for a relatively low-mass regime ($10.2\lesssim log_{10}(M/M_{\odot}\lesssim 10.7)$) the local environment modulates this dependence.\\
The analysis of ETGs also shows clear indications of mass-downsizing: from a SED-fitting analysis, it is possible to see that massive galaxies have older stellar populations, with an age difference of $\approx 3$ Gyrs within a mass range $10<log_{10}(M/M_{\odot})<10.8$, using passive evolution models with high age/$\tau$ values.
We also evaluated the formation redshift of {\it ETGs} by performing a best fit to the SED of these galaxies using different stellar population synthesis models. The lookback times evaluated for ETGs provide additional support of a downsizing scenario; from Fig. \ref{fig:fig18} it is possible to see that massive galaxies ($log_{10}(M/M_{\odot})\sim11$) formed their stars at $z\sim2$, while less massive ones ($log_{10}(M/M_{\odot})\sim10.25$) formed their stars at $z\lesssim1$, almost independently of the models of stellar population synthesis considered. Our results also agree with Thomas et al. (2009), which were based on an analysis of SDSS early-type galaxies.\\
Environment appears to have a far smaller influence than mass on galaxy properties studied. The age difference between different environments is insignificant ($\lesssim0.2$ Gyrs) and inconsistent with the difference of 2 Gyrs found by Thomas et al. (2005). Other studies found much smaller differences. For example, Bernardi et al. (2006) found differences in the $Mg_{2}-\sigma_{V}$ relation of galaxies in different environments implying that galaxies in dense environments are at most 1 Gyr older than galaxies in low density environments, Gallazzi et al. (2006) found that ETGs in dense environments are $\sim0.02$-dex older than in the field, and later analyse by Thomas et al (2007) and Thomas et al (2009) of SDSS found no dependence of scaling relation on environmental density. Moreover, we recall that the zCOSMOS survey samples a smaller range of overdensities than these previous studies, extending at most to rich groups but not clusters.
A future interesting prospect of this work is the study of the evolution of the ages and spectral features of ETGs as a function of redshift, because that analysis would also help to place constraints on cosmological parameters, such as the Hubble constant $H_{0}$ and the equation of state parameter of Dark Energy $w_{DE}$.
\begin{acknowledgements}

We acknowledge support from an INAF contract PRIN-2007/1.06.10.08 and an ASI grant ASI/COFIS/WP3110 I/026/07/0.

\end{acknowledgements}

\clearpage

%%-------------------------------------------------------------------------------

\begin{thebibliography}{}
\bibitem{Baldry} Baldry, I.K., Glazebrook, K., Brinkmann, J. et al. 2004, ApJ, 600, 681
\bibitem{Balogh99} Balogh, M., Morris, S., Yee, H. et al. 1999, ApJ, 527, 54B
\bibitem{Balogh04} Balogh, M. L., Baldry, I. K., Nichol, R. et al. 2004, ApJ, 615, 101
\bibitem{Bernardi03a} Bernardi, M., Sheth, R., Annis, J. et al. 2003a, AJ, 125, 1817
\bibitem{Bernardi03c} Bernardi, M., Sheth, R., Annis, J. et al. 2003c, AJ, 125, 1866
\bibitem{Bernardi06} Bernardi, M., Nichol, C., Sheth, R. et al. 2006, AJ, 131, 1288
\bibitem{Bertoldi} Bertoldi, F., Carilli, C., Aravena, M., Schinnerer, E. et al. 2007, ApJS, 172, 132
\bibitem{Bolzonella} Bolzonella, M., Kova\v{c}, K., Pozzetti, L. et al. 2009, A\&A submitted (arXiv:0907.0013)
\bibitem{BC} Bruzual, G., \& Charlot, S. 1993, ApJ, 405, 538
\bibitem{BC03} Bruzual, G., \& Charlot, S. 2003, MNRAS, 344, 1000
\bibitem{Buser} Buser, R. 1978, A\&A, 62, 411B
\bibitem{Capak} Capak, P., Aussel, H., Ajiki, M. et al. 2007, ApJS, 172, 99
\bibitem{Caputia} Caputi, K. I., Kova\v{c}, K., Bolzonella, M. et al 2009, ApJ, 691, 91
\bibitem{Caputib} Caputi, K. I., Lilly, S. J., Aussel, H. et al 2009, ApJ, 707, 1387
\bibitem{Cassata} Cassata, P., Guzzo, L., Franceschini, A. et al. 2007, ApJS, 172, 270
\bibitem{Cassata08} Cassata, P., Cimatti, A., Kurk, J. et al. 2008, A\&A, 483, 39
\bibitem{Chang} Chang, R., Gallazzi, A., Kauffmann, J. et al. 2006, MNRAS, 366, 717
\bibitem{Coleman} Coleman, G. D., Wu, C.-C., Weedman, D. W. 1980, ApJS, 43, 393
\bibitem{Cooper} Cooper, M.C., Coil, A.L., Gerke, B.F. et al. (arXiv:1007.1967)
\bibitem{Cucciati} Cucciati, O., Iovino, A., Kova\v{c}, K. et al 2010, A\&A submitted
\bibitem{Franzetti} Franzetti, P., Scodeggio, M., Garilli, B. et al. 2007, A\&A, 465, 711
\bibitem{Elvis} Elvis, M., Civano, F., Vignali, C. et al., 2009, ApJS, 184, 158 (arXiv:0903.2062)
\bibitem{Gallazzi} Gallazzi, A., Charlot, S., Brinchmann, J., White, S. 2006, MNRAS, 370, 1106
\bibitem{Hamilton} Hamilton, D. 1985, ApJ, 297, 371
\bibitem{Hasinger} Hasinger, G., Cappelluti, N., Brunner, H. et al. 2007, ApJS, 172, 29
\bibitem{Hogg} Hogg, D., Blanton, M., Brinchmann, J. et al. 2004, ApJ, 601, 29
\bibitem{Kauffmann} Kauffmann, G., Heckman, T. M., White, S. D. M. et al. 2003, MNRAS, 341, 33
\bibitem{Kinney} Kinney, A. L., Calzetti, D., Bohlin, R. C. et al. 1996, ApJ, 467, 38
\bibitem{Knobel} Knobel, C., Lilly, S. J., Iovino, A. et al. 2009, ApJ, 697, 1842
\bibitem{Koekemoer} Koekemoer, A. M. et al. 2007, ApJS 172, 196
\bibitem{Kovac} Kova\v{c}, K., Lilly, S. J., Cucciati, O. et al. 2009, ApJ, 708, 505 (arXiv:0903.3409)
\bibitem{Kurk} Kurk, J., Cimatti, A., Zamorani, G. et al. 2009, A\&A, 504, 331
\bibitem{Ilbert} Ilbert, O., Arnouts, S., McCracken, H. J. et al. 2006, A\&A, 457, 841
\bibitem{Iovino} Iovino, A., Cucciati, O., Scodeggio, M. et al 2010, A\&A, 509, 40(arXiv:0909.1951)
\bibitem{Lamareille} Lamareille, F., Contini, T., Le Borgne, J.-F. et al. 2006, A\&A, 448, 893
\bibitem{LeBorgne} LeBorgne, J.-F., Bruzual, G., Pell\'o, R. et al 2003, A\&A, 402, 433
\bibitem{LeFevre} Le~F\`evre, O., Saisse, M., Mancini, D. et al. 2003, SPIE, vol. 4841, p.1670
\bibitem{Lilly07} Lilly, S. J., Le F`evre, O., Renzini, A. et al. 2007, ApJS, 172, 70
\bibitem{Lilly09} Lilly, S. J., Le Brun, V., Maier, C. et al. 2009, ApJS, 184, 218
\bibitem{M05} Maraston, C. 2005, MNRAS, 362, 799
\bibitem{McCraken} McCracken, H.J., Capak, P., Salvato, M. et al. 2010, 708, 202
\bibitem{Mignoli} Mignoli, M., Zamorani, G., Scodeggio, M. et al. 2009, A\&A, 493, 39
\bibitem{Peng} Peng, Y., Lilly, S.J., Kovac, K. et al 2010, submitted (arXiv:1003.4747P)
\bibitem{Pozzetti} Pozzetti, L., Bolzonella, M., Zucca, E. et al. 2009, submitted (arXiv:0907.5416)
\bibitem{Renzini} Renzini, A. 2006, ARA\&A, 44, 141
\bibitem{Sanders} Sanders, D. B., Salvato, M., Aussel, H. et al. 2007, ApJS, 172, 86
\bibitem{Scarlataa} Scarlata, C., Carollo, C. M., Lilly, S. J. et al. 2007a, ApJS, 172, 406
\bibitem{Scarlatab} Scarlata, C., Carollo, C. M., Lilly, S. J. et al. 2007b, ApJS, 172, 494
\bibitem{Schiavon} Schiavon, R. P., Faber, S. M., Konidaris, N. et al. 2006, APJ, 651, 93
\bibitem{Schinnerer} Schinnerer, E., Smol\v{c}i\`{c}, V., Carilli, C. L. et al. 2007, ApJS, 172, 46
\bibitem{Schmidt} Schmidt, M. 1968, ApJ, 151, 393
\bibitem{Scoville} Scoville, N., Aussel, H., Brusa, M. et al. 2007, ApJS, 172, 1
\bibitem{Taniguchi} Taniguchi, Y., Scoville, N., Murayama, T. et al. 2007, ApJS, 172, 9
\bibitem{Tasca} Tasca, L., Kneib, J.P., Iovino, A. et al. 2009, A\&A, 503, 379
\bibitem{Thomas} Thomas, D., Maraston, C., Bender, R., Mendes de Oliveira, C. 2005, ApJ, 621, 673
\bibitem{Thomas07} Thomas, D., Maraston, C., Schawinski, R. et al. 2007, IAUS, 241, 546
\bibitem{Thomas09} Thomas, D., Maraston, C., Schawinski, R. et al. 2009, submitted (arXiv:0912.0259)
\bibitem{Vergani} Vergani, D., Zamorani, G., Lilly, S. J. et al 2010, A\&A, 509, 42 (arXiv:0909.1968)
\bibitem{Zamojski} Zamojski, M. A., Schiminovich, D., Rich, R. M. et al. 2007, ApJS, 172, 468
\bibitem{Zucca06} Zucca, E., Ilbert, O., Bardelli, S. et al. 2006, A\&A, 455, 879
\bibitem{Zucca09} Zucca, E., Bolzonella, M., Bardelli, S. et al. 2009, A\&A, 508, 1217
\end{thebibliography}
\end{document}